# Chemical Abundances in Clusters of Galaxies

Francesca Matteucci[1] and Brad K. Gibson[2,3]

[1]Dipartimento di Astronomia, Università di Trieste,
Via G.B. Tiepolo, 11, 34131 Trieste

[2]Department of Astrophysics, University of Oxford
Keble Road, Oxford, OX1 3RH

[3]Department of Geophysics & Astronomy,
University of British Columbia
Vancouver, B.C., Canada V6T 1Z4

**ABSTRACT** We study the origin of iron and $\alpha$-elements (O, Mg, Si) in clusters of galaxies. In particular, we discuss the [O/Fe] ratio and the iron mass-to-luminosity ratio in the intracluster medium (ICM) and their link to the chemical and dynamical evolution of elliptical and lenticular galaxies. We adopt a detailed model of galactic evolution incorporating the development of supernovae-driven galactic winds which pollute the ICM with enriched ejecta. We demonstrate *quantitatively* the crucial dependence upon the assumed stellar initial mass function in determining the evolution of the mass and abundances ratios of heavy elements in typical model ICMs. We show that completely opposite behaviours of [$\alpha$/Fe] ratios (i.e. positive versus negative ratios) can be obtained by varying the initial mass function without altering the classic assumptions regarding type Ia supernovae progenitors or their nucleosynthesis. Our results indicate that models incorporating somewhat flatter-than-Salpeter initial mass functions (i.e. $x \approx 1$, as opposed to $x = 1.35$) are preferred, provided the intracluster medium iron mass-to-luminosity ratio, preliminary [$\alpha$/Fe]>0 ASCA results, and present-day type Ia supernovae rates, are to be matched. A simple Virgo cluster simulation which adheres to these constraints shows that $\sim 70\%$ of the measured ICM iron mass has its origin in type II supernovae, with the remainder being synthesized in type Ia systems.

# 1 Introduction

In the past two decades a great deal of attention has been devoted to the study of the abundances of heavy elements (mostly iron) in the intracluster medium (ICM). From the hot X-ray emitting intergalactic gas in clusters of galaxies a universal abundance of iron of roughly 1/2 solar has been be derived (e.g. Rothenflug & Arnaud 1985).

Attempts to explain this ICM iron abundance can be traced back to the early 1970s, when for the first time the iron-emission line in the X-ray spectra of clusters of galaxies was discovered (Mitchell et al. 1976; Serlemitsos et al. 1977). Some interpreted the presence of heavy elements in the ICM as due to gas ejected from galaxies, either by means of galactic winds or ram pressure stripping (e.g. Gunn & Gott 1972; Larson & Dinerstein 1975; Vigroux 1977; Sarazin 1979; Himmes & Biermann 1980), while others suggested pregalactic objects such as population III stars as the origin (e.g. White & Rees 1978).

In the 1980s, because of the improvement in modeling the chemical evolution of elliptical galaxies, the evolution of single isotopes could be followed in space and time more accurately (Matteucci and Tornambè 1987). Since then, more detailed predictions concerning the ICM and its composition have been available. The first work of this type was by Matteucci & Vettolani (1988, hereafter MV88), followed by Pastor et al. (1989), White (1991), David, Forman & Jones (1991), Ciotti et al. (1991), Arnaud et al. (1992), Okazaki et al. (1993), Renzini et al. (1993), Mihara & Takahara (1994) and Elbaz, Arnaud & Vangioni-Flam (1995).

Here we intend to model the evolution of iron and several $\alpha$-elements (e.g. O, Mg,

and Si - elements which are synthesized from $\alpha$-particles) in the ICM of galaxy clusters. The importance of studying Fe and $\alpha$-elements in the ICM of galaxy clusters resides in the fact that these are key elements for understanding the formation and evolution of galaxies and galaxy clusters. Knowledge of the $\alpha$/Fe ratio relative to solar in the intergalactic gas can impose very strong constraints on the formation and evolution of galaxy clusters and elliptical galaxies, and in particular, can allow one to discriminate the SN type Ia and II iron contribution to the ICM. MV88 explained the observed mass of iron in the ICM of several local clusters by invoking galactic winds in the cluster ellipticals which eject their SNe I and II enriched ISM. The contribution from SNe of different types (II, Ia and Ib) was taken into account. The iron contributed by a single galaxy was then integrated over a Schechter (1976) luminosity function for a particular galaxy cluster in order to derive the total amount of iron restored to the ICM from all the cluster galaxies. This allowed them to conclude that cluster galaxies can be responsible for all of the observed amount of iron in the ICM, but that they cannot be responsible for the observed total mass of gas. They found, in fact, that the total amount of gas produced and restored by ellipticals in clusters is at most a factor of ten lower than the observed one. This fact led them to conclude, for the first time, that most of the gas in galaxy clusters has a primordial origin. The same conclusion has been reached by most of the more recent studies.

MV88 also made predictions for the expected ratio of $\alpha$-elements to iron in the ICM, showing that they should be underabundant relative to solar ratios (i.e. [$\alpha$/Fe]<0.0). This was simply a consequence of assuming that the type Ia SNe were responsible for the bulk of iron observed in the ICM, due either to the use of the Salpeter (1955) IMF or to the ubiquitous presence of late-time continuous SNe Ia-driven winds in all the models. This conclusion was in agreement with a number of parallel studies adopting detailed models for iron production (e.g. Ciotti et al. 1991; Abia, Canal & Isern 1991; Renzini et al. 1993). Subsequent to, and contrary to MV88, White (1991), David, Forman & Jones (1991), Arnaud et al. (1992) and most recently, Elbaz, Arnaud & Vangioni-Flam (1995), concluded that the iron in the ICM could be due predominantly to SNe II, and consequently, the predicted [$\alpha$/Fe] ratios should in fact be positive. In order to come to this result, it was necessary to adopt an IMF which was biased towards high-mass stars – either flatter-than-Salpeter or bimodal – which are the primary source of newly synthesized $\alpha$-elements.

An important question to ask is if elliptical galaxies enrich the ICM only through the early galactic winds and retain the gas restored from stars after the wind epoch or if all the gas restored by stars is later lost through late-time SNe type Ia-driven winds. Our earlier study (MV88) concluded that the galaxies were susceptible to the latter scenario. It is important to note that this version of our code did not include any dark matter component. The subsequent addition of this component (Matteucci 1992) makes a re-examination of the conclusions of MV88 a timely one.

At present, there are very few observations of $\alpha$-element abundances in galaxy cluster which might aid us in discriminating between SNe type Ia or type II as the prime contributors to the ICM abundances. Spectroscopic measures of elements such as Si and S are known for the Perseus cluster (Mushotzky et al. 1981) and A576 (Rothenflug et al. 1984). They indicate that the abundance of these elements as well as that of iron are roughly solar, although their level of accuracy does not allow to make strong statements about possible differences between S, Si and Fe. Concerning oxygen there are only a few of measures of the O VIII line in Virgo and Perseus (Canizares, Markert & Donahue 1988) indicating a higher than solar [O/Fe]. By far the most encouraging advances in this field have come from the recent ASCA X-ray satellite. Preliminary results from measurements of 6 clusters by Mushotzky (1994) imply [O/Fe]≈ +0.1 → +0.7 and [Si/Fe]≈ +0.0 → +0.5.

The aim of this paper is to recalculate the total masses of O, Mg, Si, Fe as well as the total gas masses ejected by elliptical and lenticular galaxies in clusters with different richness, and give new predictions for the abundances and abundance ratios in the ICM, under different assumptions about the iron ejection from galaxies. The chemical evolution model for elliptical galaxies that we intend to use is that of Matteucci (1992; 1994) which is one of the more sophisticated and self-consistent packages available. The main difference between the model used here and that of MV88 is the inclusion of dark matter in the form of a massive diffuse halo. Different stellar initial mass functions (IMFs) will be tested, as opposed to MV88 which used only the Salpeter (1955) IMF. Another improvement relative to all of the previous papers is that we calculate the blue luminosities of elliptical galaxies so that we do not have to assume a $M/L_B$ ratio *a priori*, irrespective of the assumed IMF.

In Section 2 the chemical evolution model for the elliptical galaxies is described. In Section 3 the results are presented, and in Section 4 the relevant conclusions are discussed.

## 2  The chemical evolution model

### 2.1  Basic equations

The adopted model of galactic evolution is that outlined by Matteucci & Tornambè (1987) and Matteucci (1992), where extensive descriptions can be found. The evolution of the abundances of several chemical species (H, He, C, N, O, Ne, Mg, Si and Fe) in the gas is followed, taking into account detailed nucleosynthesis from stars of all masses and SNe of types Ia, Ib, and II. We assume that ellipticals can be considered initially as homogeneous spheres of gas with luminous mass in the range $10^9 \to 2 \times 10^{12}$ M$_\odot$. A single zone interstellar medium (ISM) with instantaneous mixing of gas is assumed throughout. The adopted age for all galaxy models is $t_{\rm G} = 15$ Gyr.

The fundamental equations can be written

$$\frac{dG_i(t)}{dt} = -\psi(t)X_i(t) + \int_{M_{\rm L}}^{M_{\rm Bm}} \psi(t-\tau_m)Q_{mi}(t-\tau_m)\phi(m){\rm d}m$$
$$+ A\int_{M_{\rm Bm}}^{M_{\rm BM}} \phi(m)\left[\int_{\mu_{\rm min}}^{0.5} f(\mu)\psi(t-\tau_{m2})Q_{mi}(t-\tau_{m2}){\rm d}\mu\right]{\rm d}m$$
$$+ (1-A)\int_{M_{\rm Bm}}^{M_{\rm BM}} \psi(t-\tau_m)Q_{mi}(t-\tau_m)\phi(m){\rm d}m$$
$$+ \int_{M_{\rm BM}}^{M_{\rm U}} \psi(t-\tau_m)Q_{mi}(t-\tau_m)\phi(m){\rm d}m, \qquad (1)$$

where $G_i(t) = \rho_{\rm gas}(t)X_i(t)/\rho(0)$ is the volume gas density in the form of an element $i$ normalized to the initial total volume gas density. The quantity $X_i(t)$ represents the abundance by mass of an element $i$ and by definition the summation over all the elements present in the gas mixture is equal to unity.

The various integrals in equation 1 represent the rates at which SNe (I and II) as well as single low and intermediate mass stars and single massive stars restore their processed and unprocessed material into the ISM. In particular, the first integral refers to single stars in the mass range $M_{\rm L} = 0.8\ {\rm M}_\odot$ (the minimum mass which can restore gas to the ISM) and $M_{\rm Bm} = 3\ {\rm M}_\odot$ (the minimum mass for a binary system which can give rise to SNe Ia). The second integral represents the contribution from type Ia SNe, under the assumption that these SNe originate from white dwarfs in binary systems according to the model of Whelan & Iben (1973). The computation of the type Ia SN rate follows the prescriptions given in Greggio & Renzini (1983a,b) and Matteucci & Greggio (1986), to which we address the reader for a complete description of all the involved quantities. We only note here that the constant A appearing in equation 1 represents the fraction of systems with total mass in the appropriate range ($M_{\rm Bm} \to M_{\rm BM}$, with $M_{\rm BM} = 16\ {\rm M}_\odot$), which will give rise eventually to a SN Ia event, and is fixed *a posteriori* to ensure compatibility with the observed present-day rate of SNe Ia – i.e. $R_{\rm SNIa}({\rm present}) = (0.25 \to 0.44)h^2$ SNu , with $h = H_0/100$ km/s/Mpc (Turatto, Cappellaro & Benetti 1994)–. The mass $m_2$ is the mass of the secondary star in the binary system, which sets the clock for the explosion; the function $f(\mu)$ is the distribution function for the mass fraction of the secondary ($\mu = m_2/M_{\rm B}$) and is taken from Greggio & Renzini (1983a). The third and fourth integrals refer to single stars in the mass range of interest for type Ia SNe (i.e. $3 \to 16$ M$_\odot$) and to single massive stars (i.e. $16 \to 100$ M$_\odot$), respectively. The quantity $Q_{mi}$ represents the fraction of a star of mass $m$ which is restored to the ISM in the form of an element $i$. A description of the computational method for this quantity can be found in Ferrini et al. (1992). Finally, the sum of the four integrals in equation 1 over all the chemical elements represents the total rate at which stars restore their recycled gas to the ISM.

The star formation rate $\psi(t)$ is given by

$$\psi(t) = \nu \rho_{\rm gas}(t)/\rho(0). \qquad (2)$$

i.e. normalized to the initial total volume density. $\psi(t)$ is assumed to drop to zero at the onset of the galactic wind. The quantity $\nu$ is expressed in units of Gyr$^{-1}$ and represents the efficiency of star formation, namely the inverse of the time scale of star formation. The values adopted here for $\nu$ are, in most of the cases, the same as in Matteucci (1992). In one case we have adopted the prescription for the **inverse wind model** of Matteucci (1994). The difference between the two cases is that in Matteucci (1992) (i.e. what we shall call the **classic wind model**) one assumes that the efficiency of star formation decreases with increasing total galactic mass, as in Arimoto & Yoshii (1987), whereas Matteucci (1994) explored the scenario whereby the efficiency of star formation increases as the total galactic mass increases (similar in spirit to the SFR efficiency parametrization of Tinsley & Larson 1979) leading to a situation in which more massive galaxies experience a galactic wind before the less massive ones. As demonstrated by Matteucci (1994), this inverse wind model may play a role in accounting for the apparent trend of increasing [Mg/Fe] with $M_{\rm tot}$, as observed in the cores of giant ellipticals (Worthey, Faber & Gonzalez 1992).

The quantity $\tau_m$ represents the lifetime of a star of mass $m$, and is taken from Padovani & Matteucci (1993).

The initial mass function (IMF) by mass $\phi(m)$ is defined as

$$\phi(m) = Cm^{-x}, \qquad (3)$$

where $x = 1.35$ is the Salpeter (1955) index (this corresponds to an IMF by number $\propto m^{-(1+x)}$) and the normalization is performed in the mass range $0.1 \leq M/{\rm M}_\odot \leq 100$. We have considered three different IMF models:

i) Salpeter (1955) index $x = 1.35$ for the entire mass range,

ii) Arimoto & Yoshii (1987) index $x = 0.95$ for the entire mass range, and

iii) Kroupa, Tout & Gilmore (1993) IMF. This is a multiple power law defined as follows:

$x = +0.30$ for $0.10 \leq M/{\rm M}_\odot \leq 0.50$

$x = +1.20$ for $0.50 \leq M/{\rm M}_\odot \leq 1.00$

$x = +1.70$ for $1.00 \leq M/{\rm M}_\odot \leq 100$

The nucleosynthesis prescriptions for type II SNe (i.e. stars with $M > 10$ M$_\odot$) are taken from Woosley (1987) and Arnett (1991), those for type Ia SNe are from Nomoto et al. (1984) (their model W7), and those for single intermediate mass stars ($0.8 \leq M/{\rm M}_\odot \leq 8$) are from Renzini & Voli (1981). Although more recent yields from massive stars are now available (e.g. Woosley & Weaver 1993), the difference between the old and new yields for the elements of interest here is negligible. Finally, the type Ib SNe are included in the type II SNe, under the assumption that they originate from Wolf-Rayet stars.

## 2.2 Galactic winds

The existence of galactic winds in ellipticals was first suggested by Mathews & Baker (1971) and Larson (1974) in order to explain the apparent lack of gas in these systems and to reproduce the well known mass-metallicity relation. Although we now know, from X-ray studies of ellipticals, that these systems contain a substantial fraction of hot gas ($10^8 - 10^{11} M_\odot$; e.g. Fabbiano 1989), the existence of a wind phase at some stage of evolution of these galaxies is still required both to avoid overproducing gas and to explain the observed iron abundance in the ICM. For gas to be expelled from a galaxy the following condition should be satisfied: the thermal energy of the gas heated by SN explosions should exceed the binding energy of the gas (Larson 1974). At this point the gas present in the galaxy is swept away and the subsequent evolution is determined only by the amount of matter and energy which is restored to the ISM by the dying stellar generations. In particular, only low mass stars contribute to this evolutionary phase and, among the SNe, only SNe of type Ia (i.e. those SNe events whose progenitors have the longest lifetimes).

Therefore, in order to evaluate the time for the onset of a galactic wind we need to know the energy input from SNe and the binding energy of the gas as a function of time. The total thermal energy of the gas at time t, $E_{\rm th_{SN}}$, and the binding energy of the gas in presence of a diffuse halo of dark matter are calculated as described in Matteucci (1992). In particular, $E_{\rm th_{SN}}(t)$ is calculated by assuming that $\sim 70\%$ of the initial blast wave energy is transferred into the ISM as thermal energy by a SN remnant, if the time elapsed from the SN explosion is shorter than a SN remnant cooling time (Cox 1972). The percentage of transferred energy then decreases as a power law in time $\propto t^{-0.62}$ for times larger than the cooling time.

We do not consider the energetic input from mass loss during the thermally pulsing regime of the asymptotic giant branch for low mass stars (Vassiliadis & Wood 1993), nor envelope ejection in planetary nebula phase (Van Buren 1985) since they are not important contributors to the system thermal energy. This is attributed to the extremely small ejection velocities involved in both cases ($\gtrsim 50$ km/s), compared with those from stellar winds ($\simeq 2000$ km/s) and supernovae.

The energetic input from stellar winds in massive stars is also ignored, since for normal ellipticals it is negligible compared to the SNe thermal energy contribution, as shown by Gibson (1994). This is in contrast with a recent suggestion by Bressan, Chiosi & Fagotto (1994) who claim that the wind energy input from massive stars far exceeds that due to SNe, and as a result is the only important trigger in driving a galactic wind in ellipticals. Their results, as pointed out by Gibson, are due in part to an overestimate of the efficiency of thermalization they assumed for the stellar winds. This, coupled with an inappropriate wind phase duration (pre-SN lifetime), a neglect of the post-main sequence evolutionary status (winds are not constant in magnitude from the zero age main sequence up to the SN time, as assumed in their study), and an adopted decline in the post-wind phase residual stellar wind energy which grossly overestimates the available energy to drive a galactic wind, leads to the conclusion that stellar winds can be neglected for galaxies of mass $M \gtrsim 10^9$ M$_\odot$.

The calculation of $E_{b_{\rm gas}}(t)$ is strongly influenced by assumptions concerning the presence and distribution of dark matter. Matteucci (1992) showed that, if dark matter exists in ellipticals, realistic models (e.g. reproducing the observed metal content and present time type Ia SN rates) require it to be distributed in diffuse halos. In fact, under these conditions a negligible difference is produced in the model results, as compared to the results of standard models without dark matter We have assumed here that the ratio between dark and luminous mass is $M_{\rm dark}/M_{\rm lum} = 10$ and that the ratio between the effective radius of luminous matter and the radius of dark matter is $r_{\rm lum}/r_{\rm dark} = 0.1$.

Finally, the condition for the onset of a galactic wind is written as

$$E_{\rm th_{SN}}(t_{\rm GW}) = E_{\rm b_{gas}}(t_{\rm GW}), \tag{4}$$

where $t_{\rm GW}$ indicates the time of the occurrence of the galactic wind. We assume that the wind devoids the galaxy of all the gas present at $t_{\rm GW}$. After the wind has occurred the gas restored by the dying stars is likely to be hot, as type Ia SNe continue to inject energy in this rarefied ISM. This ISM is in fact probably responsible for the observed X-ray emission in ellipticals. So under these conditions it is justified to assume that no star formation will take place after the occurrence of a galactic wind, although we cannot exclude that other minor episodes of star formation may occur. This assumption is justified generally for normal elliptical galaxies by the fact that their photometric and chemical properties can be best explained under the assumption that they are old systems, where star formation stopped several Gyr ago (Buzzoni, Gariboldi & Mantegazza 1992). The time $t_{\rm GW}$ either increases with the galactic mass as a consequence of condition (4) and the efficiency of star formation decreasing with time, or decreases with the galactic mass if the efficiency of star formation is strongly decreasing with galactic mass, as can be seen in Table 1, and discussed further in Section 3.

The current epoch galaxy B-band luminosity was estimated using the fuel consumption theorem of Renzini & Buzzoni (1986) applied to a simple stellar population (SSP), coupled with the typical observed rest-frame (B-V)$_0$ colors from Brocato et al. (1990)

who studied the synthesis of stellar populations with different metallicities but only for two different IMFs. As in Padovani & Matteucci (1993), a multiplicative factor of 1.4 was applied to the SSP-determined $L_B$. This factor takes account of the difference between the SSP-determined blue luminosities and those calculated by Brocato et al. (1990).

## 2.3 The chemical enrichment of the ICM

Following the formalism developed in MV88, where a detailed description can be found, we will approximate the relations between the ejected masses of the heavy elements and total gas by an elliptical galaxy of final luminous mass $M_f$ by power laws:

$$M_i^{ej} = E_i M_f^{\beta_i}, \qquad (5)$$

where $M_i^{ej}$ represents the mass ejected in the form of the chemical species $i$ by a galaxy of mass $M_f$, and $E_i$ and $\beta_i$ are two constants. Once these relations are known, one can integrate over the mass function of the cluster obtained from the Schechter (1976) luminosity function. The final expression for the total mass $M_{i,\text{tot}}^{ej}$ restored from all galaxies in a cluster with mass larger than $M_f$ is

$$
\begin{aligned}
M_{i,\text{tot}}^{ej}(> M_f) &= E_i f n^* (h^2 K)^{\beta_i} 10^{-0.4\beta_i(M_B^* - 5.48)} \\
&\quad \cdot \Gamma\left[(\alpha + 1 + \beta_i), (M_f^* h^2/K) 10^{-0.4(M_B^* - 5.48)}\right],
\end{aligned} \qquad (6)
$$

where $f$ represents the fraction by number of ellipticals and lenticulars in a cluster, $n^*$ is the cluster richness, $h = H_0/100$ km/s/Mpc, $K$ is the mass-to-blue luminosity ratio $M_f/L_B$, $M_B^*$ is the absolute blue magnitude of the galaxy at the "break" of the Schechter luminosity function and $M_f^*$ its luminous final mass, $\Gamma(a, b)$ is the incomplete Eulerian $\Gamma$ function, and $\alpha$ is the slope of the luminosity function. As can be seen from inspection of equation 6, the total masses restored to the ICM from all the E/S0s in the cluster depend on several parameters, including the Hubble constant $H_0$, the mass-to-luminosity ratio $K$, and the cluster richness $n^*$ (aside: and consequently $M_f^*$, since these latter two parameters are related, as discussed in MV88).

# 3 Results

Adopting the supernovae-driven galactic wind formalism of Section 2, we present in Table 1 the predicted ejected mass of gas and various elements (Fe, O, Mg, and Si) from galaxies of initial luminous mass $M_g(0)$. We note that for each IMF we have chosen *a posteriori* the appropriate binary parameter $A$ (from equation 1) to ensure the present-day SNe Ia rate in ellipticals is compatible with the observed value (Turatto, Cappellaro & Benetti 1994). Note that the type Ia rates for the inverse wind model in column 8 of Table 1 are too low by a factor $3 \rightarrow 10$ for giant ellipticals ($M \gtrsim 10^{12}$ $M_\odot$). This was noted in Matteucci (1994) (e.g. see her Table II), and does not alter the conclusions presented in our current work.

Recall that the global impulsive galactic wind is presumed to occur once the residual thermal energy of all supernovae remnants exceeds the binding energy of the remaining interstellar gas. For the classic wind models of Table 1, the global wind occurs later in more massive galaxies, and thus more of the initial gas has become locked into stars and remnants. Only for the smaller initial masses ($\lesssim 10^{10}$ $M_\odot$) does the ratio of ejected gas to initial mass exceed a few percent. This is further illustrated by Figure 1 in which we show the evolution of the gas mass fraction as a function of time for a small subset of the models in Table 1. Models with smaller $t_{\rm GW}$ can be seen to eject a larger fraction of their ISM. We mention in passing that, as expected, the opposite behaviour is encountered when we adopt the inverse wind model outlined in Matteucci (1994) (i.e. wind epochs are inversely proportional to the initial galactic mass). In all cases, $t_{\rm GW}$ is consistently between several million years and ~2 Gyrs.

It is important to note that all of our model galaxies undergo only a global "impulsive" ejection of their ISMs at $t_{\rm GW}$. We did not encounter situations in which a continuous, post-$t_{\rm GW}$, SN Ia-driven wind developed. This is a major difference between the results of our current work, and that of its predecessor, MV88. The models in MV88 did not include any massive, diffuse, dark matter component, the result of which was the post-$t_{\rm GW}$ system binding energy was insufficient to retain the gas returned to the ISM by dying low mass stars and type Ia SNe. The deeper potential wells resulting from the massive dark halos in our current formalism cause this hot gas to be retained by the galaxy. This leads, for example, to a substantial reduction (i.e. up to an order of magnitude less) in the predicted mass of iron ejected by a given elliptical (compare our Table 1 with Table 2 of MV88). Our current models, with their absence of substantial, late-time winds is in agreement with the complementary work of Bressan, Chiosi & Fagotto (1994) and Elbaz, Arnaud & Vangioni-Flam (1995). The accumulation of low-mass ejecta is easily seen in the upturn in the post-$t_{\rm GW}$ gas mass fraction in Figure 1. It is to be expected that ram pressure stripping of some fraction of this post-$t_{\rm GW}$ (and indeed perhaps some fraction of the pre-$t_{\rm GW}$ ISM) by the surrounding dense ICM may play some part in the enrichment process (e.g. Sarazin 1979), so our tabulated results could be taken as a rough lower limit, although White (1991) does present a persuasive argument against such stripping playing a major role.

The predicted mass-to-luminosity ratios in column 9 of Table 1 should be compared to the observed ones. Typically values between 5 and 10, obtained with $H_0 = 50$ km/s/Mpc, are preferred (Bender, Burstein & Faber 1992). However, since $M/L$ scales linearly with the Hubble constant, one should take into account a factor of two uncertainty in these values. All of our models give mass-to-luminosity ratios compatible with the quoted uncertainty on the Hubble constant and show that $M/L$ is strongly dependent on the assumed IMF. Therefore, as discussed more extensively in Matteucci (1994), possible trends in the $M/L$ ratio with luminosity require a variation of the IMF from galaxy to galaxy.

Finally, the mass-weighted mean stellar iron abundance, calculated as in Matteucci (1994), for each model galaxy is given in column 10 of Table 1. Regardless of IMF, each of the classic wind models predicts an increase of the iron abundance with galactic mass and luminosity, while the Salpeter IMF inverse wind model does not. The predictions of the inverse wind model are in qualitative agreement with the data of Worthey et al. (1992) about the behaviour of Mg and Fe abundances in ellipticals. These authors, in fact, suggested that while the Mg abundance is increasing with galactic mass and luminosity (the classic age-metallicity relation) the abundance of Fe seems to have a much flatter behaviou, as indicated by the observed $Mg_2$ and $<Fe>$ indices. Matteucci (1994) discussed extensively this problem and suggested the inverse wind model as one of the possible solutions, since it predicts an increasing abundance of Mg with galactic mass and a flat behaviour for Fe. However, in this context where we are mainly interested in the total iron ejected by all ellipticals in a cluster, the use of a classic or inverse wind model is not relevant because of the negligible differences produced in the final integrated masses of the various elements. There is only a small difference in the predicted [O/Fe] ratios (see Table 4), in the sense that inverse wind models tend always to give slightly lower [O/Fe] ratios.

The evolution of the iron abundance in the ISM (i.e. $[Fe/H]_{ISM}$) for a subset of the models in Table 1 is shown in Figure 2. This illustrates graphically the separate contribution of type Ia and II SNe to the predicted mass of iron ejecta at $t_{\rm GW}$ (i.e. column 4 of Table 1). The distribution depends upon both the time of ejection, as well as the binary parameter $A$. For example, the top and bottom panels show that for the Arimoto & Yoshii IMF, type Ia SNe contribute $\sim$50% of the iron at $t_{\rm GW}$ for the most massive galaxies, and $\sim$30% for the $10^{11}$ M$_\odot$ model. For still lower mass models this drops to $\sim 10 \rightarrow 20$%. On the other hand, a Kroupa et al. IMF with its correspondingly larger value for $A$ (0.30 versus 0.05 for the Arimoto & Yoshii IMF) leads to the conclusion that virtually all of the iron ejected at $t_{\rm GW}$ originates in type Ia SNe. This holds for the lower mass models as well. In Figure 3, we show a related plot of the [O/Fe] evolution in the ISM of some of the Table 1 models. Positive values of [O/Fe] at $t_{\rm GW}$ are found only for the Arimoto & Yoshii IMF (although even the most massive models (i.e. $M \gtrsim 10^{12}$ M$_\odot$) are mildly negative). Only the $10^9$ M$_\odot$ model with the Kroupa et al. IMF leads to [O/Fe]>0, while Salpeter models with $M \gtrsim 10^{10}$ M$_\odot$ are all negative. These latter points (type Ia iron contribution at $t_{\rm GW}$ and ISM [O/Fe] at $t_{\rm GW}$) have serious repercussions for our predicted cluster ICM [O/Fe] ratios, as we shall discuss momentarily.

Using the data presented in Table 1 we have generated a series of power law fits (see equation 5) relating final galactic masses $M_{\rm f}$ to the mass of ejected species $i$ ($M_i^{\rm ej}$). The coefficients $E_i$ and $\beta_i$, as defined in Section 2.3, are listed in Table 2 for the three classic and one inverse wind model shown in Table 1. With these power-law fits, we next used equation 6 to predict the contribution of galactic winds to the gas and metal abundances for a number of galaxy cluster models.

Table 3 shows the primary input ingredients which were adopted to represent typical rich and poor clusters of galaxies. Self-consistency of the mass-to-light ratios of column 3 with those of Table 1 is an improvement over the older MV88 work where $M/L_B$ was a free parameter. The parameter $\alpha = -1.3$ is chosen as representative of the early-type luminosity function slope (e.g. Binggeli, Sandage & Tammann 1987). The values chosen for $f$, $n^*$ and $M_B^*$ are taken predominantly from MV88, where detailed references can be found. In particular, we have considered two extreme cases: specifically, rich and poor clusters. For rich clusters we take $n^* = 115$ and $f = 0.8$, while for poor clusters we use $n^* = 20$ and $f = 0.3$. In the majority of the models we have assumed $H_0 = 85$ km/s/Mpc, since this choice of $H_0$ gives the best agreement with the observed iron mass-to-light ratio. Models with $H_0 = 50$ km/s/Mpc have been also calculated but none of them provided a consistent fit to all of the observed constraints. In Table 4 we only report the results obtained with $H_0 = 50$ and Arimoto & Yoshii IMF, which is the best model for this choice of $H_0$.

Table 4 shows the predicted mass of gas and elements ejected by E/S0s into the ICM for the cluster models listed in Table 3. Column 6 contains the so-called Iron Mass-to-Light Ratio (IMLR) as defined by Renzini et al. (1993) equation 2. We have estimated the representative cluster IMLR from the ratio $M_{Fe}^{ej}/L_B$ for an $M_G^*$ galaxy (column 8 of Table 3, with the appropriate mass-to-luminosity ratio taken from Table 1). The observed IMLR of the ICM (from Renzini et al. 1993) is typically $M_{Fe}^{ICM}/L_B = 0.01 \to 0.02$, under the assumption of $h = 0.5$. As pointed out correctly by Mihara & Takahara (1994), this quantity is dependent upon the Hubble constant. In particular, $M_{Fe}^{ICM} \propto h^{-5/2}$, while $L_B \propto h^{-2}$, thus the IMLR is proportional to $h^{-1/2}$. Therefore, the range of variation of the observed IMLR, taking into account the uncertainty on the Hubble constant, is $(0.007 \to 0.014)h^{-1/2}$. Column 7 of Table 4 gives the predicted ICM [O/Fe]. It is important to recall that according to the most recent results of Mushotzky (1994) (see also the earlier work of Canizares, Markert & Donahue (1988), for example), this ratio, in the four clusters for which oxygen has been strongly detected, is [O/Fe]≈ +0.1 → +0.7. Column 8 of Table 4 shows our predicted ICM [Si/Fe] ratios for the model clusters in question. Mushotzky (1994) suggests [Si/Fe]≈ +0.0 → +0.5. As one can see, the predicted [Si/Fe] ratios are slightly lower than the [O/Fe] ratios, as also indicated by the observed values. This effect depends on the slightly higher contribution to Si than O by the SNe of type Ia, as originally discussed in Matteucci & Greggio (1986).

Taken together, all the above observational constraints, (i.e. the observed IMLRs, ICM gas [O/Fe], and present-day SNe Ia rates), compared with the results of columns 6 and 7 of Table 4 (and column 8 of Table 1), favour the wind model using the Arimoto & Yoshii (1987) IMF (i.e. $x = 0.95$). By far the strongest constraint would seem to be the ICM [O/Fe] ratio. Other than the models with an IMF biased towards high-mass stars (i.e. the $x = 0.95$ slope), all the remaining classic and inverse wind models result in [O/Fe] ratios of $-0.3 \to -0.7$. The IMLR does not appear to be as strong a constraint as [O/Fe]$_{ICM}$. For example, while most of the IMFs lead to marginally low values for IMLR, even a modest amount of ram pressure stripping of the post-$t_{GW}$ ISM would lead to substantial increases in the mass of iron being deposited to the ICM by a given galaxy due to the continual ISM iron replenishment by type Ia SNe. Of course, stripping this iron-enriched gas only drives the [O/Fe]$_{ICM}$ downwards, worsening even further this particular constraint.

The top panel of Figure 4 shows the evolution of the "poor cluster" ICM [O/Fe] ratio for both the Arimoto & Yoshii and Kroupa et al. IMFs. For the flatter IMF, we also include a breakdown of the SNe Ia and II contributions. The observational lower limit of [O/Fe]$_{ICM}$ ≳ 0.1 (Mushotzky 1994) is marked. Clearly, the Kroupa et al. IMF leads to significantly low values for the ratio. The lower panel shows only the ICM iron mass evolution as a function of time. Again, for the Arimoto & Yoshii IMF we find that approximately 70% of the iron observed in the ICM originates from type II SNe, the remainder being made up of type Ia ejecta.

In the bottom panel of Figure 4, we also show the observed range for a recognized poor cluster – the Virgo cluster – specifically, $M_{\text{Fe}}^{\text{Virgo}} = (1.6 \rightarrow 3.2) \times 10^9 h^{-5/2}$ M$_\odot$ (Okazaki et al. 1993), when h=0.85 is chosen. Again, the flatter IMF yields a predicted ICM iron mass in better agreement than the steeper Kroupa et al. IMF. Interestingly enough, the model with $H_0$=50 km/s/Mpc and Arimoto & Yoshii IMF does not predict enough ICM iron mass for the Virgo cluster but the [O/Fe] ratio is higher than in the case with $H_0$=85 km/s/Mpc, in better agreement with the preliminary results from ASCA. The reason for this behaviour of the ICM [O/Fe] ratio resides in the fact that the mass at the "break" of the Schechter luminosity function, $M_{\text{G}}^*$, decreases with the assumed Hubble constant, as is shown in Table 3. Lower mass elliptical galaxies, in fact, develop a galactic wind earlier than more massive galaxies and the composition of their ejected material contains more O than Fe relative to the more massive objects. This effect would be opposite in the inverse wind model. Unfortunately, at the moment we cannot clearly distinguish between the classic and the inverse wind models for ellipticals. However, what matters here is that irrespective of the assumed $H_0$ one can obtain [O/Fe]> 0 only if a flat IMF is chosen, and no post-$t_{\text{GW}}$ winds or ram pressure stripping is occurring.

Finally, not surprisingly, as in MV88, the predicted mass of ejected gas for all the models in Table 4 (i.e. see column 1) is lower than that measured (e.g. column 3 of Table 1 in MV88) by one to two orders of magnitude, implying the necessary presence of a primordial gas component.

# 4  Conclusions

We have presented new calculations of the amount of Fe and $\alpha$-elements which can be restored to the ICM by ellipticals and lenticulars in poor and rich clusters of galaxies. To do that we have adopted a detailed and self-consistent model of galactic evolution where the energy input from SNe of all types (II, Ia and Ib), as well as dark matter, are taken into account. The evolution of ellipticals is, in fact, mainly determined by the interplay between the energy input from SNe and the depth of the potential well, which influences the occurrence of galactic winds. Such an occurrence, in turn, determines the amount of heavy elements which remain locked into stars and that which can be lost to enrich the ICM. We have explored different IMF prescriptions and derived self-consistent $M/L_{\rm B}$ ratios via use of the fuel consumption theorem. Along with the presence of dark matter, these are all improvements upon our earlier work (MV88). The amount of gas per galaxy available to be ejected has then been integrated over a Schechter luminosity function to obtain the total amount of gas and metals contributed by all galaxies in the clusters.

Our main conclusions can be summarized as follows:

i) the IMF is a most crucial parameter in determining dynamical, chemical and photometric galactic evolution, as it influences the timescale for the occurrence of a galactic wind, as well as the contribution of type II versus type Ia SNe to the enriched ejecta. Our analysis shows that only a flat IMF ($x \approx 1$) can meet simultaneously the requirements of having a high [$\alpha$/Fe] ratio in the ICM, the appropriate IMLR, the observed SNe Ia rates, for the underlying cluster ellipticals. The pollution of the ICM occurs only through early galactic winds, since the mass restored by stars after the wind phase has completed remains bound to the galaxy, similar to what was found by Elbaz, Arnaud & Vangioni-Flam (1995). On the other hand, wind models with Salpeter-like and Kroupa et al. (1993) IMFs (i.e.

for IMFs which are more suitable for the solar vicinity region) may reproduce the IMLR and SN Ia rates, but the resultant negative ICM [O/Fe] ratios is at odds with the most recent ASCA observations. Additional ram pressure stripping of cluster ellipticals in their post-$t_{\rm GW}$ phase seems to be excluded by our models, in agreement with the arguments of White (1991). This further strengthens the case for [$\alpha$/Fe] ratios representing a useful diagnostic tool in interpreting galaxy and galaxy cluster evolution.

ii) The abundance of Fe, predicted for the stellar gas restored after the early wind phase is over, is always solar or oversolar in the studied models regardless of the assumed IMF. This seems reasonable, since studies of stellar populations in ellipticals (Weiss et al. 1995) indicate that the average abundance in the dominant stellar population of these galaxies is roughly solar. Therefore, none of our models would be consistent with Fe abundances in the X-ray halos of ellipticals which are substantially subsolar, unless dilution from the ICM has occurred. Recent X-ray studies of Fe abundances in elliptical galaxies seem to give contradictory information (Serlemitsos et al. 1993; Forman et al. 1994) so that no firm conclusions can yet be drawn.

iii) All the models require the presence of type Ia SNe and give present time type Ia supernova rates consistent with the observed ones. Our best model for the cluster ICM abundances (Arimoto & Yoshii IMF) required a binary parameter $A = 0.05$ in equation 1 (approximately a factor of two lower than that needed for solar neighbourhood models - e.g. Matteucci & Greggio 1986). In such a scenario, while type II SNe are the dominant contributor to the ICM iron abundance – for a typical cluster, approximately 3/4 of the iron originates there – we note that type Ia SNe still provide the non-negligible remaining $\sim 25\%$.

iv) Finally, we confirm the previous result of MV88 regarding the total amount of

unenriched gas that galaxies can contribute to the ICM. As before, we find that the galaxies in clusters, both poor and rich, cannot account for all of the hydrogen gas observed. One is forced to conclude that the bulk of hydrogen in galaxy clusters has a cosmological origin.

*Acknowledgements* FM wishes to thank L. Norci, P. Padovani and S. Pellegrini for many useful discussions during her stay at E.S.O.

**FIGURE CAPTIONS**

**Fig. 1.** Predicted behaviour of the gas mass fraction for galaxies of different initial luminous mass and different IMF. The minimum in the curves indicates the time of occurrence of the galactic wind. After the wind is over the amount of gas starts increasing again due to the dying stars.

**Fig. 2.** Predicted behaviour of the [Fe/H] abundance in the ISM of galaxies of different initial luminous mass and IMF. The contributions of supernovae of different types are shown.

**Fig. 3.** Predicted behaviour of the [O/Fe] ratio in the ISM of galaxies of different mass and IMF.

**Fig. 4. Upper panel**: Predicted behaviour of the [O/Fe] ratio in the ICM of a poor cluster (e.g. Virgo) as a function of the IMF of the member galaxies. The different contributions from supernovae of different type are also shown. The arrow marks the observational lower limit of Mushotzky (1994). **Lower panel**: Predicted behaviour of the iron mass in the ICM for the same cluster as in the upper panel. The contributions of supernovae of different types are also shown. The arrow marks the observed range for the iron mass in the ICM of the Virgo cluster (Okazaki et al. 1993) when $H_0$=85 km/s/Mpc is assumed.